\documentclass[a4paper,11pt]{article}
\pdfoutput=0 % if your are submitting a pdflatex (i.e. if you have
\usepackage{jcappub} % for details on the use of the package, please
\usepackage{natbib}
\usepackage{dcolumn}
\usepackage{graphicx}
\usepackage{color}
\usepackage[usenames,dvipsnames]{xcolor}
\usepackage[normalem]{ulem}
\usepackage{multirow}
\usepackage{ulem}

\newcommand{\apj}{Astrophys.~J.}
\newcommand{\mnras}{Mon.~Not.~R.~Astron.~Soc.}

\newcommand{\be}{\begin{equation}}
\newcommand{\ee}{\end{equation}}
\newcommand{\bea}{\begin{eqnarray}}
\newcommand{\eea}{\end{eqnarray}}

\newcommand{\E}{\mathcal{E}}

\title{Effects of anisotropy on gravitational infall in galaxy clusters using an exact general relativistic model}

\author[a]{M. A. Troxel}
\author[a]{Austin Peel}
\author[a]{Mustapha Ishak}

\affiliation[a]{Department of Physics, The University of Texas at Dallas, \\Richardson, TX 75083, USA}
\emailAdd{troxel@utdallas.edu}
\emailAdd{austin.peel@utdallas.edu}
\emailAdd{mishak@utdallas.edu}

\abstract{
We study the effects and implications of anisotropies at the scale of galaxy clusters by building an exact general relativistic model of a cluster using the inhomogeneous and anisotropic Szekeres metric. The model is built  {from} a modified Navarro-Frenk-White (NFW) density profile. We compare this to a corresponding spherically symmetric structure in the Lema\^itre-Tolman (LT) model and quantify the impact of introducing varying levels of anisotropy. We examine two physical measures of gravitational infall -- the growth rate of density and the velocity of the source dust in the model. We introduce a generalization of the LT dust velocity profile for the Szekeres metric and demonstrate its consistency with the growth rate of density. We find that the growth rate of density in one substructure increases by 0.5\%, 1.5\%, and 3.75\% for 5\%, 10\%, and 15\% levels of introduced anisotropy, which is measured as the fractional displaced mass relative to the spherically symmetric case. The infall velocity of the dust is found to increase by 2.5, 10, and 20 km s$^{-1}$ (0.5\%, 2\%, and 4.5\%), respectively, for the same three levels of anisotropy. This response to the anisotropy in a structure is found to be strongly nonlinear with respect to the strength of anisotropy. These relative velocities correspond to an equivalent increase in the total mass of the spherically symmetric structure of 1\%, 3.8\%, and 8.4\%, indicating that not accounting for the presence of anisotropic mass distributions in cluster models can strongly bias the determination of physical properties like the total mass. }

\begin{document}


\maketitle

\section{Introduction}\label{intro}

The precision and availability of observational data in astrophysics is rapidly improving, and the degree to which we can constrain our models of the universe on both small and cosmological scales will grow dramatically over the next two decades. One question, which must be answered as a result of this, is the degree to which our interpretations can be biased by the limitations imposed by assumptions inherent in the models that we work to constrain. On cosmological scales, for example, the concordance Lambda Cold Dark Matter ($\Lambda$CDM) model of cosmology has proven robust in describing the evolution of the universe on large scales. However, $\Lambda$CDM is based upon the Friedmann-Lema\^itre-Robertson-Walker (FLRW) metric in general relativity, which is globally isotropic (and thus globally homogeneous). On galaxy and cluster scales, models of dark matter halo density like the popular Navarro-Frenk-White (NFW) profile \cite{nfw97} are often assumed to dominate the dynamics of systems, but are also isotropic. In light of the wealth of new and precise data soon to be available to us, relaxing these assumptions of homogeneity or isotropy and questioning the robustness of our interpretations will be a key test for cosmology and astrophysics.

The effects of allowing for inhomogeneity and anisotropy have begun to be explored in cosmology by using more general, exact solutions to Einstein's field equations to model an inhomogeneous and anisotropic universe \cite{SKMHH2003,Krasinski1997}. One such solution is the Szekeres models \cite{Szekeres1,Szekeres2}, which have an irrotational dust source with no symmetries (i.e., no Killing vectors) \cite{Bonnor&Tomimura1976}. The Szekeres models have been compared to supernovae Ia observations of the expansion history of the universe \cite{Ishaketal2008,Bolejko&Celerier2010,Nwankwoetal2011}, the growth history of large-scale structure \cite{Ishak&Peel2012,Peel2012}, as well as to some Cosmic Microwave Background constraints \cite{BolejkoCMB,Buckley}. The collapse of anisotropic structures in the Szekeres models has been initially explored by \cite{barrowsilk}, but for a class of the models which we do not consider here that generalize the Kantowski-Sachs universes \cite{kantowskisachs}. The Szekeres models represent a more realistic, but still exact, description of the lumpy universe we observe \cite{Bonnoretal1977,Ellis&VanElst1998}. Studies of the growth of large-scale structure with the Szekeres models have already shown that the presence of nonlinear evolution in an inhomogeneous and anisotropic model can allow for stronger growth \cite{Bolejko2006} and significant departures from standard interpretations of cosmological information, including the precise amount of matter (baryonic and dark) needed to match the growth history of large-scale structure \cite{Peel2012}. 

Including the effects of anisotropy in modeling dark matter halos on galaxy and cluster scales has also been investigated, primarily through the exploration of triaxial halo models \cite{Jing2002,Oguri2003}. It is expected that about 10\% of halo mass is present in substructure \cite{Tormen1998}, and efforts to model analytically halo substructure have also been made through the halo model \cite{Sheth2003,Giocoli2010} and by considering more realistic models with dynamic friction \cite{Oguri2004}. Anisotropies like the triaxiality of a dark matter halo can substantially bias measurements of mass and concentration. Determinations of mass through gravitational lensing can be affected by up to 50\% in cases of significant elongation along the line of sight \cite{Corless2007}, while determinations of mass and concentration are impacted at the 5\% level in weak lensing simulation catalogs of mock galaxy clusters \cite{Bahe2012}.

We combine these two approaches in this paper and use the Szekeres models to explore how including substructure anisotropies on single cluster scales in an exact general relativistic model can affect observables connected to the rate of gravitational infall. We generalize velocity measurements used in studying the Lema\^itre-Tolman (LT) models \cite{Lemaitre1933,tolman1934}, which are inhomogeneous but isotropic about a single point, to the Szekeres models, and explore the impact of anisotropy through volume and thick shell averages of the growth rate of density in the structure and the infall velocity of the dust. These calculations serve both to quantify the effect of anisotropies in kinematic observables and structure formation on the cluster scale, but also extend naturally our previous work in studying the growth of structure in the Szekeres models to a single realistic galaxy cluster.

The paper is structured as follows. In section \ref{model}, we discuss how to build an exact model of a cluster of galaxies using the Szekeres models. We describe the Szekeres model in Sec. \ref{szekeres}, and quantify the properties of the model, how it incorporates anisotropies and how this compares to the isotropic (wrt to the center) LT model in Sec. \ref{modeldef}. We introduce several measures of the strength of gravitational infall within a structure in the Szekeres models in Sec. \ref{velocity} and review how they have been used in the past. In Sec. \ref{rhot}, we describe the results of thick shell and volume averaging for the growth of density within the structure, and in section \ref{dustvel} we discuss results for the infall velocity. section \ref{summary} provides a summary of these effects and how they might impact physical measurements of properties of the cluster. Finally, we conclude in section \ref{conclusion} and discuss the implications of our results. Units are chosen throughout the paper such that $c=G=1$, unless otherwise noted. 

\section{Modeling a cluster of galaxies using the Szekeres metric}\label{model}

In order to explore the effects of anisotropy on individual structures on the galaxy cluster scale in general relativity, we must first build an exact model to represent the cluster. Clusters of galaxies are typically modeled by a spherically symmetric (isotropic) dark matter halo, which is assumed to dominate the dynamics of the system. The density profile of such halos is often described by the Navarro-Frenk-White (NFW) profile \cite{nfw97},
\be
\rho^{\textrm{NFW}}(r)=\frac{\delta_c \rho_c}{\frac{r}{r_s}(1+\frac{r}{r_s})^2},\label{eq:nfw}
\ee
where $r_s$ is a scale radius, $\rho_c$ is the critical density of the universe, and the characteristic overdensity is given by
\be
\delta_c=\frac{200}{3}\frac{c^3}{\ln(1+c)-\frac{c}{1+c}},
\ee
which is a function of the concentration of the cluster $c=r_{200}/r_s$. $r_{200}$ is the radius at which the average interior density is $200 \rho_c$.  {The radii used in Eq. (\ref{eq:nfw}) (and in Eq. (\ref{eq:nfwt}) below) are not necessarily the coordinate $r$ in a relativistic model, where coordinate and physical distances are often not equivalent. Our use of $r$ will become clear, however, in section \ref{modeldef}, where we identify the $r$ coordinate with a physical distance measurement at $t_0$ when we specify our structure. Then $r$ corresponds exactly to the angular diameter distance in the spherically symmetric case used to initially define the model.} Though we are interested in the effects of introducing anisotropies into the structure in an exact general relativistic model, the success of the NFW profile makes it a natural starting point for describing a cluster of galaxies using the Szekeres metric.

\subsection{The Szekeres models}\label{szekeres}

Our cluster model will be built from the Szekeres metric, which is an exact solution to Einstein's field equations with an irrotational dust source \cite{Szekeres1,Szekeres2}. The Szekeres models are inhomogeneous and anisotropic, which make them ideal for modeling general asymmetric structures. We use the Lema\^itre-Tolman form of the metric, which can be written as a generalization of the LT metric in synchronous and comoving coordinates as \cite{Hellaby1996}
\be
ds^2=-dt^2+\frac{(\Phi,_{r}-\Phi \E,_{r}/\E)^2}{\epsilon-k(r)}dr^2+\frac{\Phi^2}{\E^2}(dp^2+dq^2),\label{eq:metric}
\ee
where $,_{\alpha}$ represents partial differentiation with respect to the coordinate $\alpha$. The Szekeres models fall into two classes depending on the coordinate dependence of the metric functions. We use the more general Class I metric, with all metric functions having an $r$-dependence. 

The geometry of the metric is governed by $k(r)$, which determines the geometry and evolution of the spatial sections and can change sign as a function of $r$, and by $\epsilon=0,\pm 1$, which determines the geometry of the $(p,q)$ 2-surfaces. In this work, we will limit our discussion to the quasi-spherical case ($\epsilon=1$). When $\epsilon=1$, the $(p,q)$ 2-surfaces are spheres with $\Phi=\Phi(t,r)$ as their areal radius. $\Phi(t,r)$ is defined by 
\be
(\Phi,_{t})^{2}=-k+\frac{2M}{\Phi}+\frac{\Lambda}{3}\Phi^2,\label{eq:phi}
\ee
where $M=M(r)$ represents the total active gravitational mass within a sphere of constant $r$. We will choose $\Lambda=0$, because we are interested only in dynamics that occur within the cluster of galaxies, where the impact of a cosmological constant can be neglected. Equation (\ref{eq:phi}) has the same form as the Friedmann equation, but where each surface of constant $r$ evolves independently of the others.

The function $\E=\E(r,p,q)$ in Eq. (\ref{eq:metric}) is
\be	
\E(r,p,q)=\frac{S(r)}{2}\left[\left(\frac{p-P(r)}{S(r)}\right)^2+\left(\frac{q-Q(r)}{S(r)}\right)^2+\epsilon\right],\label{eq:E}
\ee
and $S$, $P$, and $Q$ describe the stereographic projection from the $p$ and $q$ plane onto the unit sphere such that
\be	
\frac{(p-P(r),q-Q(r))}{S(r)}=\frac{(\cos(\phi),\sin(\phi))}{\tan(\theta/2)}.\label{eq:transform}
\ee
When $\E=\E(p,q)$ ($S$, $P$, and $Q$ are constants), Eq. (\ref{eq:metric}) is just the LT metric, and the spheres of constant $t$ and $r$ are concentric about the origin. In the Szekeres metric, the $r$-dependence of $\E$ acts to offset these spheres relative to the LT model through the contribution of $\E,_{r}/\E$ to the metric. 

The density in the Szekeres metric is given by
\be	
\kappa\rho(t,r,p,q)=\frac{2(M,_{r}-M\E,_{r}/\E)}{\Phi^2(\Phi,_{r}-\Phi\E,_{r}/\E)},\label{eq:density}
\ee
where $\kappa=8\pi$ in units where $c=G=1$. The density of an isotropic structure depends only on the total mass function $M(r)$ and areal radius $\Phi(t,r)$ (and thus the coordinates $t$ \& $r$), and coincides with the corresponding LT structure where $\E,_{r}/\E=0$. Substructure in the form of anisotropic perturbations on this corresponding LT structure are then due to the influence of $\E,_{r}/\E\ne0$, which introduces a dipole contribution to the density at a given $r$ \cite{Szekeres2}. This means that along any surface of constant $r$ where $\E,_{r}/\E\ne0$, there exists a single density peak which decreases monotonically to a density minimum located on the opposite side of the 2-sphere. If $\E,_{r}/\E=0$ at any $r$, the density is constant along that 2-sphere.

\subsection{Model construction and properties}\label{modeldef}

 {Our cluster model will be fully defined by the Szekeres metric functions, chosen to create a realistic cluster density profile that quickly becomes homogeneous and spherically symmetric outside of the structure in order to facilitate possible association with an FLRW limit as background model. }Integrating Eq. (\ref{eq:phi}) with $\Lambda=0$, we find
\be
t-t_B(r)=\int_0^{\Phi}\frac{d\Phi'}{\sqrt{-k(r)+\frac{2 M(r)}{\Phi'}}},\label{eq:phiint}
\ee
where $t_B(r)$ is the Bang time function, giving the local time of the Big Bang at some $r$. The function $t_B(r)$ is generally arbitrary when defining a model, though in a homogeneous model, $t_B(r)=\textrm{const}$.

There are then a total of six arbitrary functions necessary to fully define the Szekeres metric: $k(r)$, $M(r)$, $S(r)$, $P(r)$, $Q(r)$, and $t_B(r)$. The metric is invariant under a transformation of the coordinate $r$, so we can reduce this number to five by redefining the $r$-coordinate as $\tilde{r}=f(r)$,  {where $f(r)$ is fixed by a physical condition condition as we, for example, do in the first bulleted item below}. Our model is then defined in the following general process. Our model is then defined in the following general process:

\begin{itemize}
\item We choose the $r$-coordinate such that $\tilde{r}=\Phi(t_0,r)$, where $t_0$ is the current age of the universe. We choose $t_0=13.7$ Gyr such that it is similar to the observed age of the universe in $\Lambda$CDM \cite{agenote}. For simplicity, we will drop the tilde in references to $r$ that follow.  {This allows us to idenfity $\Phi(t,r)$ as the $r$ coordinate at the time $t_0$ that the model is specified, which constrains the curvature $k(r)$ to be fully defined by the mass function $M(r)$ and this new $r$ coordinate (see bullet point 4 below).}
\item Using Eq. (\ref{eq:density}), we set $\E,_{r}=0$ and calculate $M(r)$ from the isotropic density distribution of a corresponding LT structure at time $t_0$.  {This expression is further simplified by our choice $\Phi(t_0,r)\equiv r$ such that $\Phi,_{r}(t_0,r)=1$.} The density distribution used is based upon the NFW profile and is discussed further in section \ref{rho}.
\item We construct a Bang time function $t_B(r)$, which is discussed further in section \ref{tb}.
\item Using the solutions of Eq. (\ref{eq:phiint}) discussed by \cite{krasinski2002,hellaby2006,walters2012}, we can calculate $k(r)$ once $M(r)$, and $t_B(r)$ are known at $t_0$.  {The choice $\Phi(t_0,r)\equiv r$ defines the upper limit of the integral.}
\item Finally, we construct the functions $S(r)$, $P(r)$, and $Q(r)$, which are discussed further in section \ref{anisotropic}. These functions contribute to $\E,_{r}/\E$ and determine the anisotropy of the structure.
\end{itemize}

\begin{figure}
\centering
\includegraphics[width=.5\columnwidth]{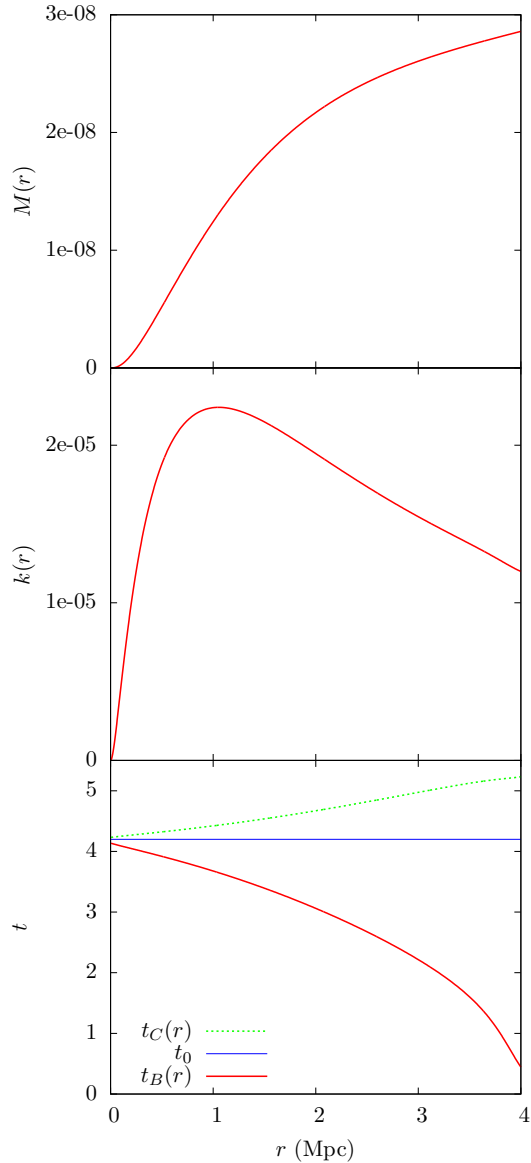}
%\vskip .75cm
\caption{\label{fig:mk}
 {Top panel: The mass function $M(r)$ from Eq. (\ref{eq:m}) for the cluster model described by the modified NFW density profile in Eq. (\ref{eq:ltdensity}). Middle panel: The resulting curvature $k(r)$ of the model, found by solving Eq. (\ref{eq:phiint}). The curvature is everywhere elliptic ($k(r)>0$) for $r>0$. Bottom panel: The Bang and Crunch times, $t_B(r)$ and $t_C(r)$, for the cluster model, with the current age of the universe $t_0=4.2$ indicated in appropriate units ($c=G=1$). Both the Bang and Crunch times asymptote to constant values at large $r$, as required in an FLRW space.}}
\end{figure}

\subsubsection{Density profile}\label{rho}

The mass function $M(r)$ of our cluster model is computed from an isotropic (LT) density profile. While the function $M$ and the density profile that initially define it are spherically symmetric, the final Szekeres density in Eq. (\ref{eq:density}) will be modified by the function $\E,_{r}/\E$ to be anisotropic. The initial isotropic density profile used is based on a modified version of the NFW profile. The NFW profile in Eq. (\ref{eq:nfw}) has two undesirable features: 1) its density diverges at small $r$, which will violate the conditions of regularity at the origin for the Szekeres metric; 2) its enclosed mass diverges at large $r$, which makes matching to an FLRW space with homogeneous density difficult. To resolve these problems, we introduce a maximum density at very small $r$, and a truncation radius $r_t$ following Baltz et al.\cite{baltz2009}
\be
\rho^{\textrm{BMO}}(r)=\frac{\delta_c \rho_c}{\left(\epsilon_c+\frac{r}{r_s}\right)(1+\frac{r}{r_s})^2}\left(\frac{r_t^2}{r^2+r_t^2}\right)^2.\label{eq:nfwt}
\ee

Our cluster model has a concentration $c=3$, $r_{200}=1.75$ Mpc, $r_t=3$ Mpc and $\epsilon_c=0.1$. As part of ensuring our metric approaches an FLRW form at large $r$ outside of the cluster, we add a constant background FLRW density with $\Omega_m=1$ \cite{lcdm} such that our final LT density profile at $t_0$ becomes
\be
\rho^{\textrm{LT}}(r)=\rho^{\textrm{BMO}}(r)+\rho_c.\label{eq:ltdensity}
\ee
Because of the introduction of the truncation radius and peak central density, the actual $r_{200}$ which results from this density profile is $1.54$ Mpc.  {Our interest in this work is limited to dynamics within the model cluster and not of the universe as a whole. Our choice of the homogeneous FLRW background density at $t_0$ does not assume that all such clusters should be described by our model simultaneously, which would imply a universe with average density well beyond critical. However, in such a treatment the total mass of overdense clusters could be offset by a sufficient number of underdense regions such that the average density remains consistent with a background $\Omega_m=1$. Here we instead simply include this FLRW limit to demonstrate that such a cluster can evolve in a universe without shell-crossing singularities and for future ease of associating such a cluster model with some background FLRW spacetime.}

 {From Eq. (\ref{eq:density})}, the mass function $M(r)$ is then defined  {at $t_0$} as the integral of
\be
M,_{r}(r)=\kappa r^2\rho_{\textrm{LT}}(r),\label{eq:m}
\ee
as discussed in section \ref{modeldef},  {where we have simplified the expression in Eq. (\ref{eq:m}) using our choice $\Phi(t_0,r)\equiv r$ such that $\Phi,_{r}(t_0,r)=1$}. The mass and curvature functions for this model are shown in the top and middle panels of figure \ref{fig:mk}. We will use the LT density in Eq. (\ref{eq:ltdensity}) as the reference density for an isotropic structure that shares the mass function $M(r)$ with our Szekeres cluster model. In section \ref{anisotropic}, we describe how the resulting Szekeres density profile which uses $M(r)$ in Eq. (\ref{eq:density}) is modified by $\E,_{r}/\E$ to be anisotropic.

\subsubsection{Bang and collapse times}\label{tb}

Two major limitations in using the Szekeres metric to model cluster scale structures at the current age of the universe is the lack of pressure or rotation in the source fluid. Collapsing structures thus have no means by which to delay or prevent direct re-collapse into a singularity. For example, the time between maximum expansion and collapse to a singularity of a Szekeres cluster like that described in section \ref{rho} is approximately 0.25 Gyrs at $r=0$. Thus to avoid a large central singularity forming in the cluster, we must either limit the age of the universe to less than 0.5 Gyrs or include a nonzero $t_B(r)$, the latter of which is preferable in our case to preserve a large $t_0$ outside of the structure.

We use a nonzero $t_B(r)$ primarily as an artificial counter to the absence of pressure and rotation in the collapse of the structure.  {This extends the lifetime of the structure so that the exterior region asymptotically approaches an FLRW model that has evolved to an age similar to the observed age of the universe.} The Bang time function
\be
t_B(r)=\left(t_0-\frac{3\pi r^{3/2}}{4\sqrt{2M(r)}}\right)\left(\frac{\pi+2\arctan{(16-4 r)}}{\pi+2\arctan{(16)}}\right)^2
\ee
is chosen such that outside an $r$ approximately 1 Mpc beyond $r_t$, $t_B(r)$ quickly approaches a near-zero constant value, as required for an asymptotically FLRW exterior. Within this $r$ at $t_0$, the structure is approximately midway through its collapsing phase. This is so that we avoid the region in time near the singularity where pressure should be dominant. The Bang time and associated Crunch time, $t_C(r)=t_B(r)+2\pi M(r)/k(r)^{3/2}$, are shown in the bottom panel of figure \ref{fig:mk}. The current age of the unviverse, $t_0$, is also shown for comparison.

\begin{figure}
\centering
\includegraphics[width=.5\columnwidth]{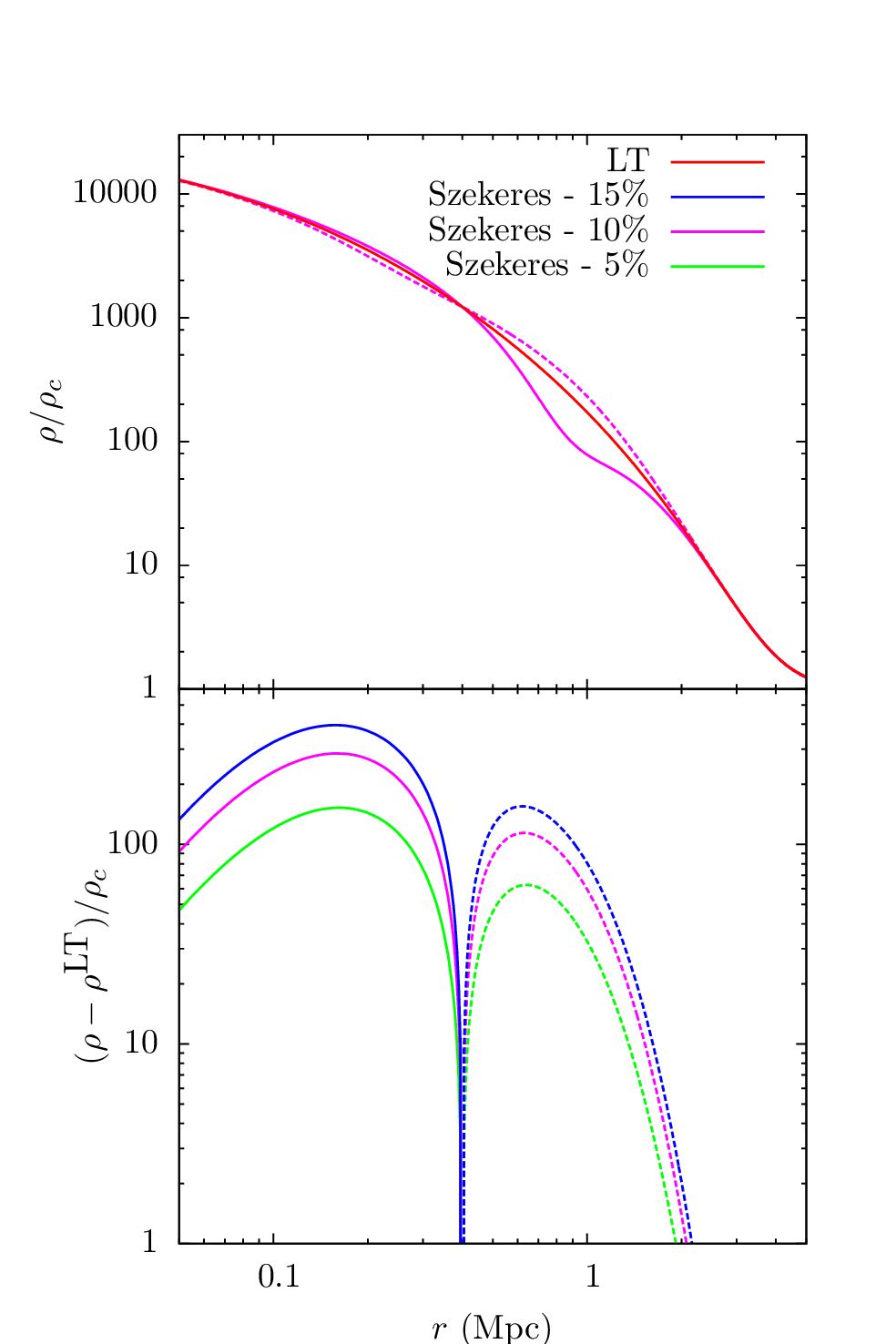}
%\vskip .75cm
\caption{\label{fig:rho}
Top panel: The density profile $\rho^{\textrm{LT}}(r)$ from Eq. (\ref{eq:ltdensity}) of the isotropic LT structure and $\rho(r)$ from Eq. (\ref{eq:density}) of the $\delta\rho_{|\mathcal{D}}/\rho^{\textrm{LT}}_{|\mathcal{D}}=10\%$ Szekeres anisotropic model. Bottom panel: The difference $\rho-\rho^{\textrm{LT}}$ for the overdense substructures in each of the Szekeres models. The Szekeres curves are measured through the directions of maximum $|\rho-\rho^{\textrm{LT}}|$. Solid lines look through the $\phi=\pi/2$ direction and dashed lines look through the $\phi=-\pi/2$ direction, as defined by Eq. (\ref{eq:transform}). In both cases, $\theta=\pi/2$. Densities are plotted in units of $\rho_c$.}
\end{figure}

\subsubsection{Anisotropic substructure}\label{anisotropic}

The functions $S(r)$, $P(r)$, and $Q(r)$, which form the metric function $E(r,p,q)$ in Eq. (\ref{eq:E}), control the anisotropic nature of the structure, but must also be chosen with several physical requirements in mind. The first is not a necessity of the metric, but rather our choice that the metric can become spherically symmetric at small enough $r$ near to the origin and at large enough $r$ sufficiently outside of the structure, as defined by $r_t$.  {Outside of $r_t$ this is to allow the space to take an FLRW limit. A sufficient condition on $E(r,p,q)$ is that $S,_{r}=0$, $P,_{r}=0$, and $Q,_{r}=0$ at any $r$ that must be spherically symmetric. This symmetry at large $r$, combined with the constant density and $t_B$ outside of the structure that was imposed in Secs. \ref{rho} \& \ref{tb}, ensures that the Szekeres metric obtains naturally an FLRW form outside of the structure. That is, the metric is spherically symmetric and $\Phi(t,r)=a(t)r$, $k(r)=k_0r^2$, $M=M_0r^3$, and $t_B(r)=\textrm{const}$.}

Second, we require that our model be free of singularities. We have ensured that we do not reach the collapse singularity by our choice of $t_B$, but we also require that
\be
\Phi,_{r}-\Phi \frac{\E,_{r}}{\E}\ne 0,\label{eq:shellcrossing}
\ee
when $k(r)\ne 1$, in order to avoid any shell-crossing singularities with an unphysical divergence in the density. Unlike the LT metric, where shells can only cross at all values of $p$ and $q$ simultaneously for a given $r$, the Szekeres metric can have shell crossings occur at a single point.  {The condition in Eq. (\ref{eq:shellcrossing}) can be shown to result in a set of conditions on the functions $M$, $k$, $S$, $P$, $Q$, and $t_B$ \cite{hellaby2002}, which our functions also satisfy.}

The final requirement on $S$, $P$, and $Q$ comes from the anisotropies we are seeking to create in our structure. We limit ourselves to a simple, but not trivial, example of anisotropy in the structure by introducing a pair of overdense/underdense regions at different $r$ values. The position of the overdense peaks in $r$ can be chosen by considering the extreme value of $\E,_{r}/\E$, which is given by \cite{hellaby2002}
\be
\left(\frac{\E,_{r}}{\E}\right)_{\textrm{ext}}=\mp\frac{\sqrt{S,_{r}^2+P,_{r}^2+Q,_{r}^2}}{S}.\label{eq:extreme}
\ee
This value is related to the strength of the Szekeres dipole contribution to the density at that $r$.

By choosing $S(r)=\textrm{const.}$ and $P(r)=0$, the right side of Eq. (\ref{eq:extreme}) simplifies to $\mp|Q,_{r}|/S$, where the dipole maxima/minima now coincide with the $r$ at which $|Q,_{r}|$ is locally maximum, and $S$ acts simply to scale the magnitude of the dipole contribution. We choose the function $Q$ (in units of Mpc) as
\be
Q(r)=27 e^{-5r}r^2,\label{eq:q}
\ee
which has a functional form that satisfies the requirement that the metric become spherically symmetric at very small and large $r$ ($Q,_{r}=0)$.  {Beyond $r=5$ Mpc, for example, $|Q,_{r}|<10^{-8}$. This degree of symmetry, along with the chosen $M(r)$ and $t_B(r)$, ensure that the metric obtains an FLRW limit beyond $r=50$ Mpc to a numerical accuracy of less than $10^{-8}$ in $\Phi=a(t)r$. This is precisely true at $t_0$, and at any times near $t_0$ where we are interested in studying the dynamics of the structure, the evolution of the space beyond $r=50$ Mpc will be numerically indistinguishable from an FLRW space to any measurable degree of accuracy.}

$Q(r)$ is then scaled appropriately to produce the desired double dipole anisotropic structure in the cluster model, while preventing the occurrence of shell-crossing singularities in the structure's past. $Q,_{r}=0$ at $r=0.4$ Mpc, which is the boundary between the two anisotropic regions. Since $S$ and $P$ are constant, this particular 2-sphere is entirely LT-like (spherically symmetric) and has a constant density. This feature will appear clearly throughout our results, where the physical differences from the LT reference structure become consistently zero at $r=0.4$ Mpc (and at very small or large $r$).

The choice of the value of $S$ can then be used to control the strength of the anisotropy in the cluster. We will define the strength of anisotropy through the total fractional displaced mass of the Szekeres structure relative to the LT reference density, defined as the volume integral $\delta\rho_{|\mathcal{D}}$ (see Eq. (\ref{eq:volint})) bounded by the 2-sphere at $r=2 r_{200}$ of the quantity
\be
\delta\rho=|\rho-\rho^{\textrm{LT}}|
\ee
at $t_0$. Since about $10\%$ of the halo mass is expected to be in substructure \cite{Tormen1998}, we choose anisotropies representing a total displacement of mass $\delta\rho_{|\mathcal{D}}/\rho^{\textrm{LT}}_{|\mathcal{D}}$ of $5\%$, $10\%$, and $15\%$, which correspond to $S=4.87$, $2.43$, and $1.65$ Mpc, respectively. The resulting density for each $S$ through the directions of largest overdensities and underdensities is shown in figure \ref{fig:rho}. The anisotropies represented by these $S$ values may be considered conservative, as we are merely shifting mass in the underlying isotropic NFW halo with the Szekeres dipole and not adding substructure mass on top of it.

\section{Measures of infall strength}\label{velocity}

The strength of gravitational infall is directly connected to the rate of structure growth in a model. The growth of large-scale structure in the Szekeres models was recently examined in \cite{Ishak&Peel2012,Peel2012}, where it was found that the growth rate is stronger in the nonlinear Szekeres models compared to the usual growth of linear perturbations in $\Lambda$CDM. In this work, we concentrate instead on a single structure and aim to measure two quantities directly related to the kinematic properties inside of the structure, and thus to the growth of the structure {: the growth of density and the infall velocity of the source dust.}

\begin{figure}
\centering
\includegraphics[width=.5\columnwidth]{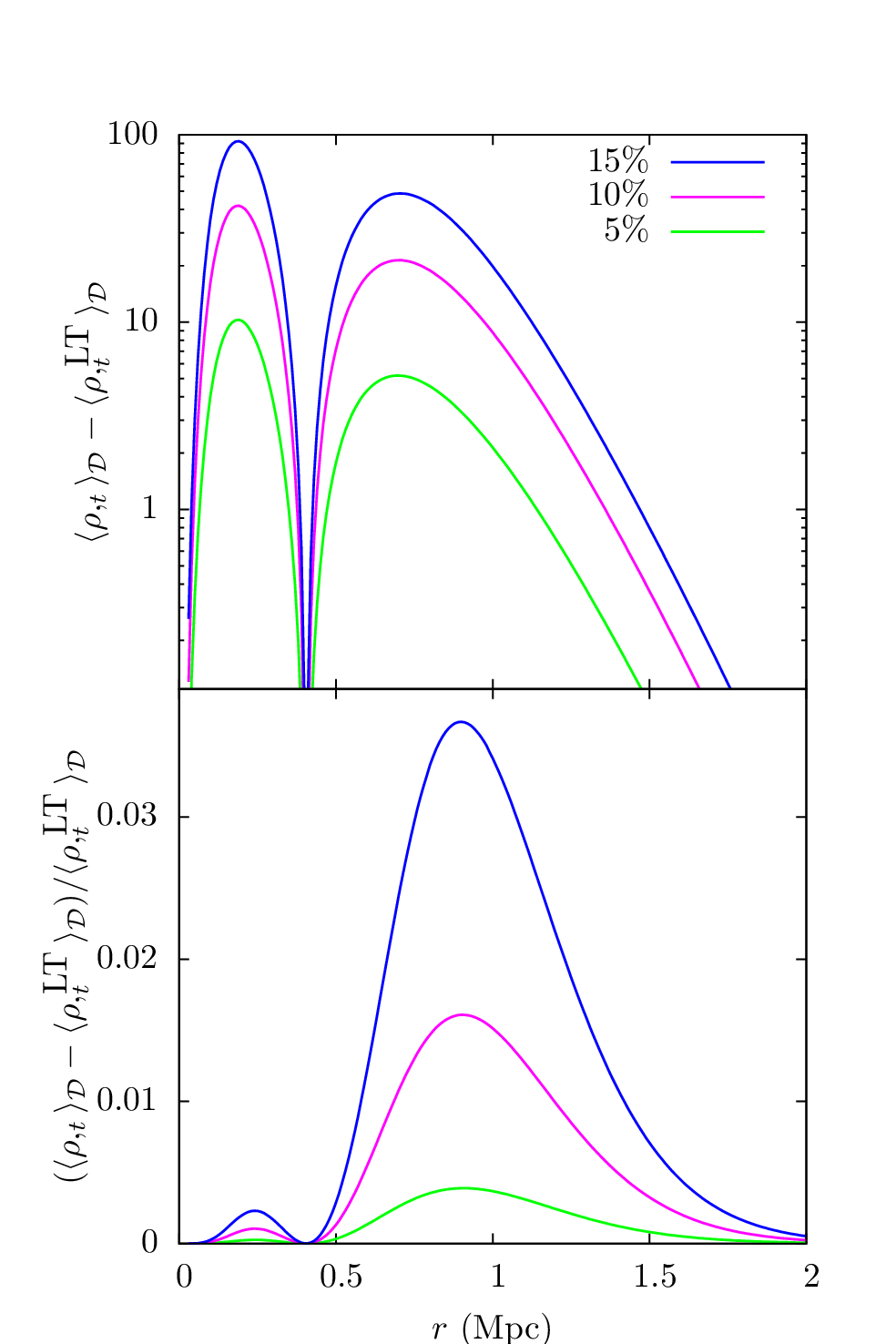}
%\vskip .75cm
\caption{\label{fig:rhot}
Top panel: The difference between the Szekeres models and the LT reference model in the growth rate of the density averaged over shells with thickness 10kpc. The Szekeres growth is stronger on average over shells within anisotropic regions. Bottom panel: The fractional difference in the average growth rate of density. The values of $r$ at which peaks in the difference of $\langle \rho,_{t}\rangle_{\mathcal{D}}$ from the reference model occur coincide with the positions of greatest anisotropy in the density. The effect on $\langle \rho,_{t}\rangle_{\mathcal{D}}$ of increasing the anisotropic displacement of mass is clearly nonlinear. Doubling $\delta\rho_{|\mathcal{D}}$ from $5\%$ of $\rho^{\textrm{LT}}_{|\mathcal{D}}$ to $10\%$ causes the difference in $\langle \rho,_{t}\rangle_{\mathcal{D}}$ to increase by a factor of 4, while tripling it to $15\%$ causes the difference in $\langle \rho,_{t}\rangle_{\mathcal{D}}$ to increase by a factor of nearly 10.}
\end{figure}

\subsection{Growth rate of density}\label{rhot}

The first of these is the growth rate of the density of the structure, which can be calculated analytically from the metric functions. The growth of structure on cosmological scales is typically explored via the density contrast 
\be	
\delta_m=\frac{\rho-\langle\rho\rangle}{\langle\rho\rangle},
\ee
which in $\Lambda$CDM is limited to a linear perturbation on the average background density of the model. In Szekeres models, $\delta_m$ can have an exact interpretation as a difference in density relative to either a global average density (Class 2 models) or a quasi-local average density (Class 1 models), which is both nonlinear and not limited to being small in magnitude. The rate of change of this density contrast is then related to model parameters and can constrain the cosmology. Because we are more interested in the kinematic differences between isotropic and anisotropic models, we adopt instead the more direct rate of change of the density itself.

The time derivative of the density in Eq. (\ref{eq:density}) is
\be	
\rho,_{t}=-\rho\left(2\frac{\Phi,_{t}}{\Phi}+\frac{H,_{t}}{H}\right),\label{eq:rhot}
\ee
where 
\be	
H\equiv \Phi,_{r}-\Phi\E,_{r}/\E.
\ee
In a collapsing region near $t_0$ for our model, $\Phi,_{t}<0$ and $\Phi,_{tr}<0$. The first term of Eq. (\ref{eq:rhot}) then contributes positively to $\rho,_{t}$ and the sign of the second term depends on the properties of $\E,_{r}/\E$. Thus we expect $\rho,_{t}$ to change sign at different $(r,p,q)$ in a general Szekeres model, but to necessarily be positive at $r$ where $\E,_{r}/\E$ become negligible (the model becomes LT-like) in a collapsing region. It is not clear a priori whether there should be a net effect in $\rho,_{t}$ on the total structure when we introduce anisotropies via a nonzero $\E,_{r}/\E$. 

To explore this, we will consider the volume average of $\rho,_{t}$ over some domain $\mathcal{D}$ of the space-time. For a full explanation of the volume averaging, see Appendix \ref{app1}. Such averages have been used in the past to demonstrate, for example, that the Szekeres dipole component $\E,_{r}/\E$ has no averaged contribution to the volume acceleration or the deceleration parameter over a volume domain interior to some $r$ \cite{bolejko2009}, and to construct a description of the dynamics of the Szekeres models through the use of `quasi-local' averaged scalars \cite{sussman2012}. We will instead consider averages over shells of finite thickness in order to examine the average properties of the cluster as a function of radius. We choose a domain $\mathcal{D}$ as in Appendix \ref{app1} such that the volume is contained between two surfaces of constant $r$ with $r-r_0=10$ kpc. 

Analytically, $\langle\rho,_{t}\rangle_{\mathcal{D}}$ can be written from Eqs. (\ref{eq:rhot}) \& (\ref{eq:volavgsz}) as
\bea
\langle \rho,_{t}\rangle_{\mathcal{D}}&=&-\frac{1}{V_{\mathcal{D}}}\int_{r_0}^{r}dr\frac{\Phi^2}{\sqrt{1-k}}\int\sin\theta d\theta\int d\phi\label{eq:volavgrhot}\\
&\times&\frac{2}{\Phi^2}\left(M,_{r}-M\frac{\E,_{r}}{\E}\right)\left(2\frac{\Phi,_{t}}{\Phi}+\frac{H,_{t}}{H}\right).\nonumber
\eea
An analytic solution to the integrals over the $(\theta,\phi)$ coordinates in Eq. (\ref{eq:volavgrhot}) is not straightforward. However, it can be shown that the Szekeres contribution to $\langle \rho,_{t}\rangle_{\mathcal{D}}$ persists in at least some cases by considering the limit where $\E,_{r}/\E\ll\Phi,_{r}/\Phi$ and expanding Eq. (\ref{eq:volavgrhot}) to apply the argument given at the end of Appendix \ref{app1}. This is lengthy, and we do not reproduce it here. Instead, we evaluate $\langle \rho,_{t}\rangle_{\mathcal{D}}$ numerically for the cluster model discussed in section \ref{model} and show the resulting averages in figure \ref{fig:rhot} compared to those for the corresponding LT structure. 

In the LT structure, $\rho,_{t}^{\textrm{LT}}(t_0,r)$ is independent of $(p,q)$. In the Szekeres cluster models, $\rho,_{t}(t_0,r,p,q)$ varies for some $r_0$, stronger than the LT growth in the overdense regions of the structure (relative to the LT density), and weaker than the LT growth in underdense regions. The average over a shell of thickness 10 kpc, however, is nonzero and stronger for the Szekeres models in anisotropic regions, as shown in the top panel of figure \ref{fig:rhot}. The fractional change is shown in the bottom panel of figure \ref{fig:rhot}. The position of the peaks of greatest change in $\rho,_{t}$ for the Szekeres models coincide with the peaks of greatest anisotropy in the structure. The effect on $\langle \rho,_{t}\rangle_{\mathcal{D}}$ of increasing the anisotropic displacement of mass is clearly nonlinear. Doubling $\delta\rho_{|\mathcal{D}}$ from $5\%$ of $\rho^{\textrm{LT}}_{|\mathcal{D}}$ to $10\%$ causes the difference in $\langle \rho,_{t}\rangle_{\mathcal{D}}$ to increase by a factor of 4, while tripling it to $15\%$ causes the difference in $\langle \rho,_{t}\rangle_{\mathcal{D}}$ to increase by a factor of nearly 10.

\begin{figure}
\centering
\includegraphics[width=.5\columnwidth]{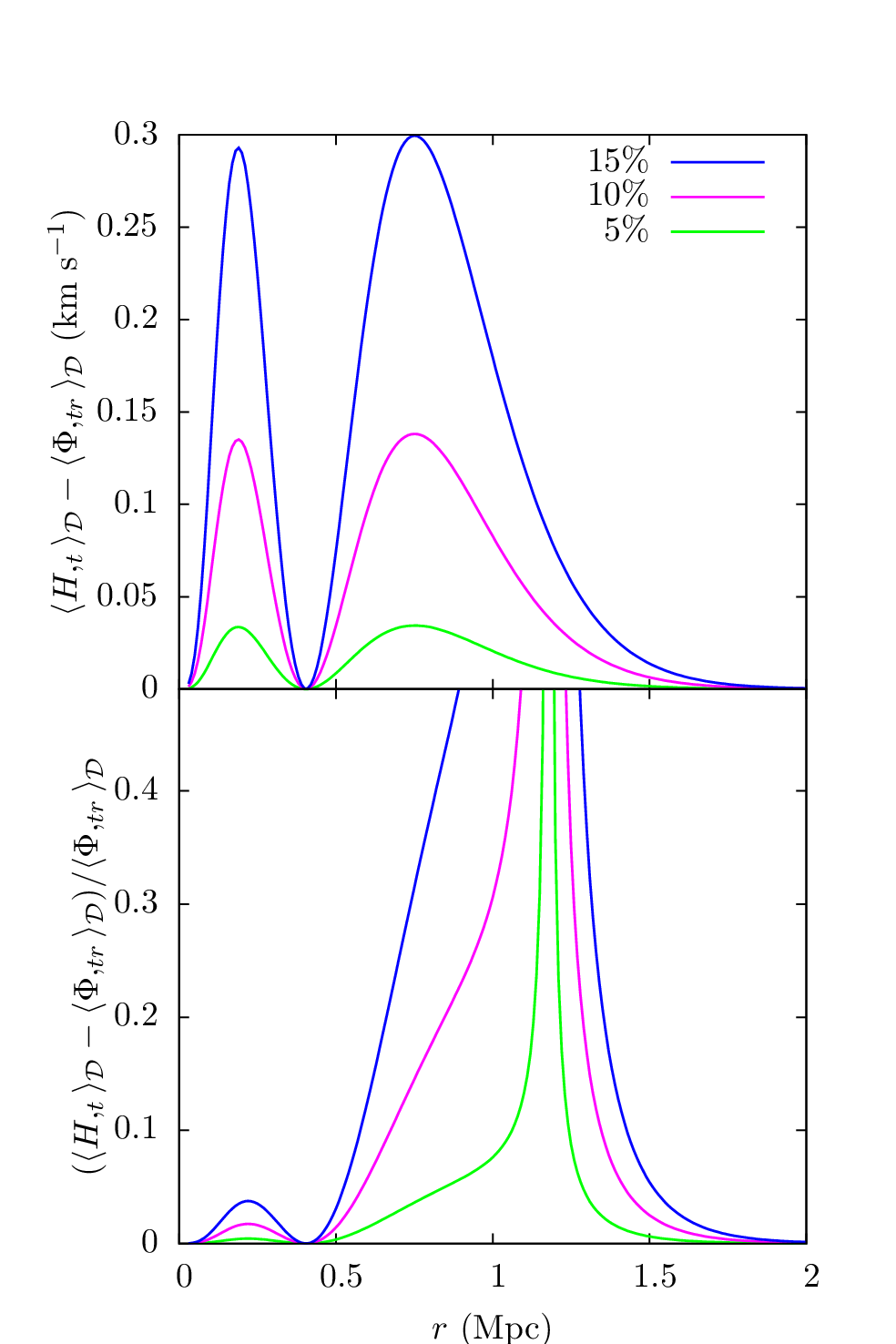}
%\vskip .75cm
\caption{\label{fig:dustvel}
Top panel: The difference between the Szekeres models and the LT reference model in the average velocity of dust $\langle H,_{t}\rangle_{\mathcal{D}}$ between shells with separations of 10kpc in km s${}^{-1}$, where $H\equiv \Phi,_{t}-\Phi \E,_{r}/\E$. The average Szekeres velocity tends to be more strongly toward neighboring shells in anisotropic regions than the LT, meaning the dust tends to cluster together more strongly in anisotropic regions. This agrees with the stronger growth rate of density in these regions seen in figure \ref{fig:rhot}, and the peaks of both measures coincide at the same $r$ values. Bottom panel: The fractional difference in this average relative velocity. The fractional difference diverges artificially around $r=1.2$ Mpc due to the sign of the velocity changing (the shells beginning to move apart from one another at higher $r$) at different $r$ values in the Szekeres and LT models. The Szekeres models continue to have velocities toward neighboring shells at higher $r$ than the LT. The effect on $\langle H,_{t}\rangle_{\mathcal{D}}$ of increasing the anisotropic displacement of mass is clearly nonlinear and follows a similar trend to that of $\langle \rho,_{t}\rangle_{\mathcal{D}}$ in figure \ref{fig:rhot}.}
\end{figure}

\begin{figure}
\centering
\includegraphics[width=.5\columnwidth]{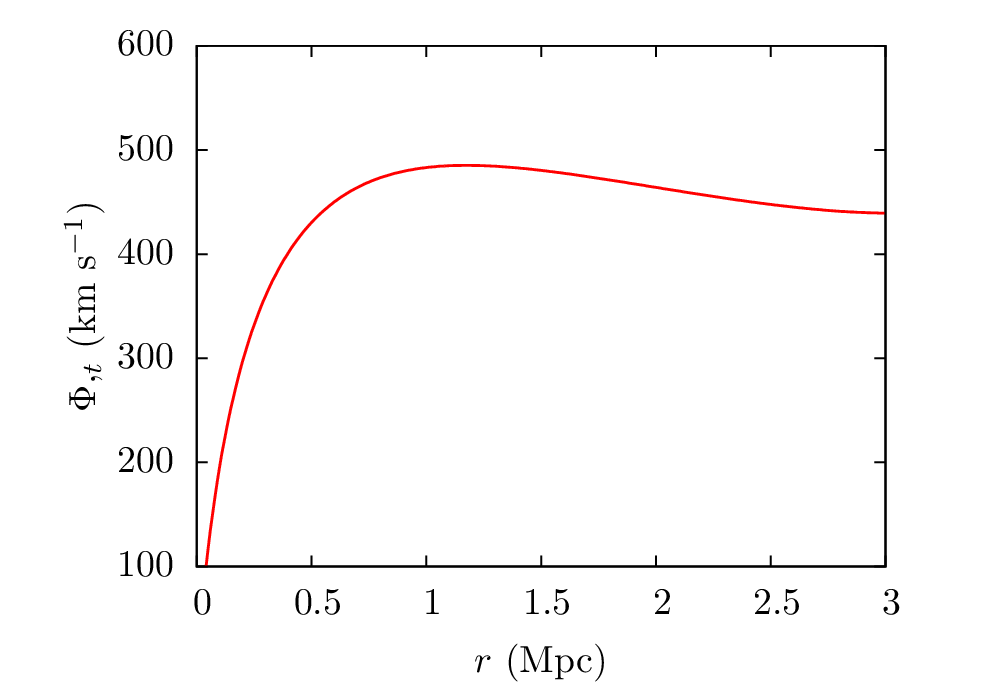}
%\vskip .75cm
\caption{\label{fig:dustvel0}
The radial infall velocity of dust in the LT reference model, as measured by $\Phi,_{t}$. The velocity follows the typical profile of a dark matter halo, reaching a peak velocity of nearly 500 km s${}^{-1}$, which is consistent with the typical galaxy cluster velocity dispersion distribution \cite{munari2013}.}
\end{figure}

\begin{figure}
\centering
\includegraphics[width=.5\columnwidth]{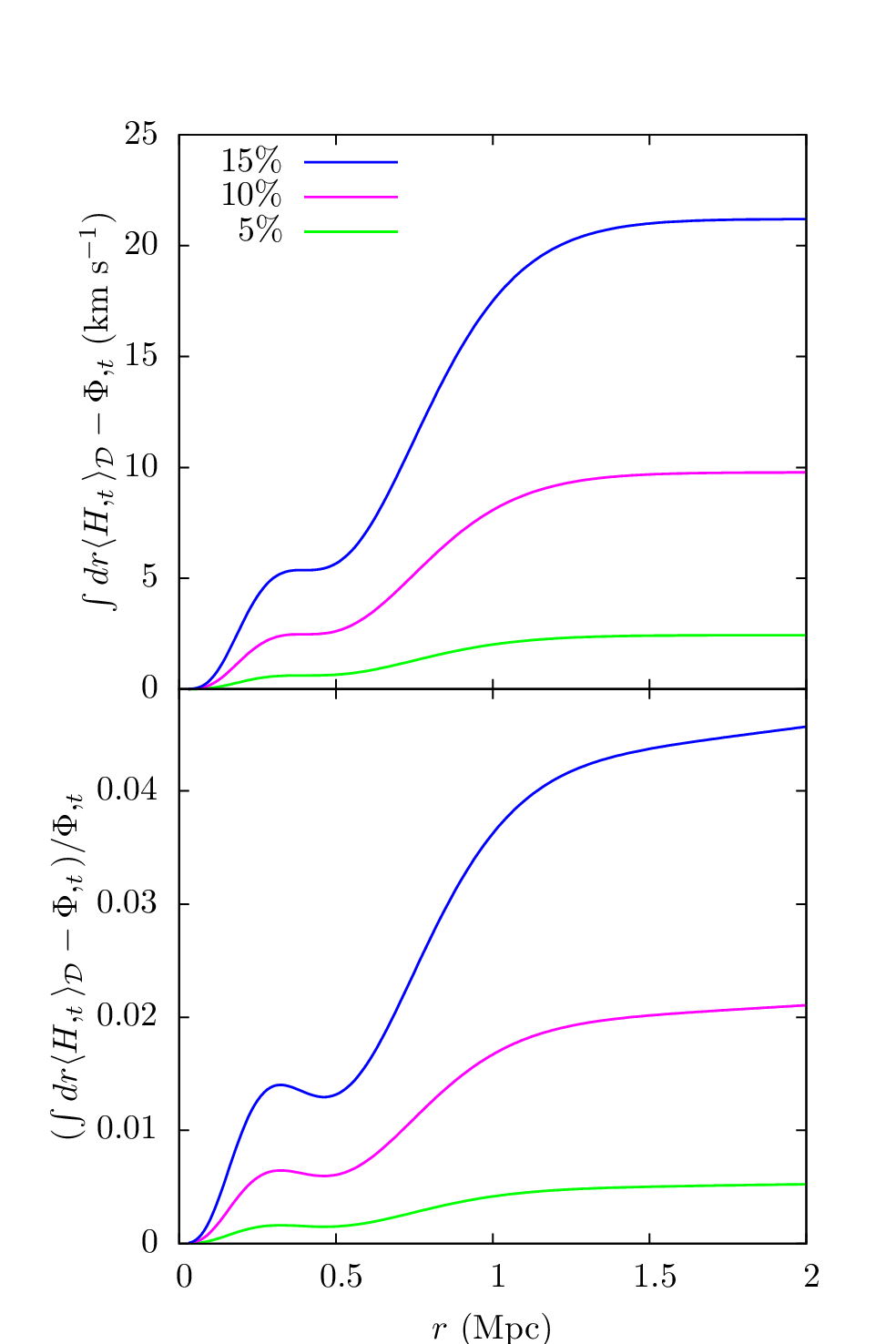}
%\vskip .75cm
\caption{\label{fig:dustvel2}
Top panel: The difference in the average infall velocity of dust between the Szekeres models ($\int dr\langle H,_{t}\rangle_{\mathcal{D}}$) and the LT reference model ($\Phi,_{t}$), where $H\equiv \Phi,_{t}-\Phi \E,_{r}/\E$. The average Szekeres infall velocity tends to grow more strongly in anisotropic regions than the LT, due to the dust clustering together more strongly in anisotropic regions (see figure \ref{fig:dustvel}). This agrees with the stronger growth rate of density in these regions seen in figure \ref{fig:rhot}. The cumulative effect of the anisotropies act to increase the average infall velocity by $2.5$, $10$, and $20$ km s${}^{-1}$, for $5\%$, $10\%$, and $15\%$ anisotropies. Bottom panel: The fractional difference in this average radial velocity, which ranges between $0.5\%$ and $4.5\%$ of the LT radial velocity. The effect on the infall velocity of increasing the anisotropic displacement of mass is clearly nonlinear and follows a similar trend to that of $\langle \rho,_{t}\rangle_{\mathcal{D}}$ in figure \ref{fig:rhot} and $\langle H,_{t}\rangle_{\mathcal{D}}$ in figure \ref{fig:dustvel}.}
\end{figure}

\subsection{Velocity of the source dust}\label{dustvel}

A second measure of the gravitational infall strength in the galaxy cluster that we consider is the velocity of the comoving source dust in the model. Quantifying this velocity is not immediately straightforward, because our coordinates are comoving, and we have defined no static metric for the galaxy cluster. However, the time evolution of the areal radius $\Phi$ has been used to quantify the velocity of the source dust in LT models (for example \cite{krasinski2002,hellaby2006}), where surfaces of constant $r$, each representing a shell of the comoving dust, are an angular diameter distance $D_A=\Phi(t,r)$ from the origin. In this sense, the quantity $\Phi,_{t}$ represents a physical radial velocity 
\be	
\Phi,_{t}=\frac{\partial D_A}{\partial t}
\ee
of the dust.

In the Szekeres models, we will consider a similar measure of velocity for the dust, which is a generalization of the LT case and takes into account the Szekeres anisotropy. Instead of simply using the area distance
\be	
D_A=\Phi=\int dr \,\Phi,_{r},
\ee
since the spheres represented by $\Phi$ are offset from the origin in the Szekeres metric, we will instead consider the rate of change of the corresponding quantity
\be	
\ell_{SZ}=\int dr \,H=\int dr \,(\Phi,_{r}-\Phi\frac{\E,_{r}}{\E})\label{eq:dsz},
\ee
which represents the analogous path length from a shell of constant $r$ to the origin in a Szekeres model. Unlike $\Phi(t,r)$, this new measure varies as a function of $(t_0,r,p,q)$. The corresponding velocity is then just 
\be	
\frac{\partial \ell_{SZ}}{\partial t}=\int dr \,H,_{t}=\int dr \,(\Phi,_{tr}-\Phi,_{t}\frac{\E,_{r}}{\E}).\label{eq:dtdsz}
\ee

We also consider a measure of velocity to sample the clustering strength at different $r$. Instead of measuring the total distance to the shell from the origin, we can instead consider the separation of adjacent shells at some $(t_0,r,p,q)$. In the LT case, this separation is just $\Phi,_{r}$. In the Szekeres models, it is $H$. Quantifying how these values change over time gives a measurement of the local velocity of the dust relative to a neighboring shell and will show how strongly it is clustering together as a function of $r$. This can be represented analytically using Eqs. (\ref{eq:volavgsz}) \& (\ref{eq:s2}) as
\bea	
\langle H,_{t}\rangle_{\mathcal{D}}&=&\frac{1}{V_{\mathcal{D}}}\int_{r_0}^{r}dr\frac{\Phi^2}{\sqrt{1-k}}\int\sin\theta d\theta\int d\phi \nonumber\\
&\times&(\Phi,_{r}-\Phi\E,_{r}/\E)(\Phi,_{tr}-\Phi,_{t}\E,_{r}/\E)\nonumber\\
&=&\frac{4\pi}{V_{\mathcal{D}}}\int_{r_0}^{r}dr\frac{\Phi^2}{\sqrt{1-k}}\label{eq:volavgvel}\\
&\times&(\Phi,_{tr}\Phi,_{r}-\Phi,_{t}\Phi\frac{S,_{r}^2+P,_{r}^2+Q,_{r}^2}{3 S^2}).\nonumber
\eea
In our cluster model, $S,_{r}=P,_{r}=0$ so that the LT contribution ($\Phi,_{tr}\Phi,_{r}$) is simply modified by a term which scales like $(Q,_{r}/S)^2$. In the collapsing region of the model, both the LT term and the Szekeres modification are negative, causing the Szekeres anisotropy to increase the clustering strength of the dust by causing neighboring shells in anisotropic regions to approach each other faster on average.

Evaluating Eq. (\ref{eq:volavgvel}) numerically as in section \ref{rhot} using shells of thickness 10 kpc, we find a result that agrees well with this analysis. This is shown in figure \ref{fig:dustvel} relative to the LT clustering. The average Szekeres velocity tends to be more strongly toward neighboring shells in anisotropic regions than the LT model, meaning the dust tends to cluster together more strongly in anisotropic regions. This agrees with the stronger growth rate of density in these regions seen in figure \ref{fig:rhot}, and the peaks of both measures coincide at the same $r$ values. The fractional difference shown in the bottom panel of figure \ref{fig:dustvel} diverges artificially around $r=1.2$ Mpc due to the sign of the velocity changing (the shells beginning to move apart from one another at higher $r$) at different $r$ values in the Szekeres and LT models. The Szekeres models continue to have velocities toward neighboring shells at higher $r$ than the LT, consistent with a stronger total growth of density in the anisotropic structure. The effect on $\langle H,_{t}\rangle_{\mathcal{D}}$ of increasing the anisotropic displacement of mass is clearly nonlinear and follows a similar trend to that of $\langle \rho,_{t}\rangle_{\mathcal{D}}$ in figure \ref{fig:rhot}.

We also calculate the cumulative effect of the clustering by integrating this average quantity $\langle H,_{t}\rangle_{\mathcal{D}}$ over $r$ as in the non-averaged Eq. (\ref{eq:dtdsz}), which gives the average velocity of the shell at $r$ toward the origin, a measure which is equivalent to $\Phi,_{t}$ in the LT metric. This total velocity of the dust toward the origin at $r$ is shown in figure \ref{fig:dustvel0} for the LT reference model. The infall velocity follows the typical velocity dispersion profile of a dark matter halo, reaching a peak velocity of nearly 500 km s${}^{-1}$, which is consistent with the typical galaxy cluster velocity dispersion distribution \cite{munari2013}.

Figure \ref{fig:dustvel2} shows the infall velocity toward the origin of the Szekeres models relative to the LT infall velocity shown in figure \ref{fig:dustvel0}. The average Szekeres infall velocity tends to grow more strongly in anisotropic regions than the LT, due to the dust clustering together more strongly there (see figure \ref{fig:dustvel}). This agrees with the stronger growth rate of density in these regions seen in figure \ref{fig:rhot}. The cumulative effect of the anisotropies acts to increase the average infall velocity by $2.5$, $10$, and $20$ km s${}^{-1}$, for $5\%$, $10\%$, and $15\%$ anisotropies. The effect on the infall velocity of increasing the anisotropic displacement of mass is clearly nonlinear and follows a similar trend to that of $\langle \rho,_{t}\rangle_{\mathcal{D}}$ in figure \ref{fig:rhot} and $\langle H,_{t}\rangle_{\mathcal{D}}$ in figure \ref{fig:dustvel}.

\begin{table}
\center
\begin{tabular}{ c c c }
\hline\hline
 ~~$S(r)$~~ & ~~Anisot. Mass Frac.~~ & LT Mass Frac. \\
 \hline
4.87 & 5\% & 1.0\% \\
2.43 & 10\% & 3.8\% \\
1.65 & 15\% & 8.4\% \\
\hline\hline
\end{tabular}
\caption{The relationship between the Szekeres function $S(r)$, the displaced mass fraction $\delta\rho_{|\mathcal{D}}/\rho^{\textrm{LT}}_{|\mathcal{D}}$ due to anisotropy in a structure as discussed in section \ref{anisotropic}, and the equivalent increase in the total mass due to a scaling of the isotropic density $\rho^{\textrm{LT}}$. The LT mass fraction is determined such that the increase in isotropic mass produces an equivalent increase in infall velocity at $r=2r_{200}$ as for the corresponding displaced anisotropic mass, as discussed in section \ref{summary}.}
\label{table1}
\end{table}

\subsection{Impacts of anisotropy on physical measurements}\label{summary}

Together, these measures of velocity and the rate of density growth agree well and provide a consistent picture of the growth being stronger in anisotropic regions of a structure. This coincides with a similar increase in the clustering velocity of the dust and causes a net increase in the average velocity of the dust toward the center of the structure (the origin of the model) in the Szekeres anisotropic cluster models. This velocity increase has a direct impact on kinematic measurements of real structures, which are based primarily on assumptions of spherical symmetry in the halo. We have shown in figure \ref{fig:dustvel2}, for example, that even a simple dipole anisotropy of 10\%, like that shown in figure \ref{fig:rho}, can have an impact of approximately $10$ km s${}^{-1}$ in the infall velocity profile, which is about $2\%$ of the total radial infall velocity of a corresponding spherical LT structure, and can grow up to nearly $5\%$ for a 15\% anisotropy.

We can quantify the real impact such anisotropies might have on determinations of properties like the total cluster mass by comparing the velocity changes due to the anisotropies to those due to simply increasing the total mass by scaling $\rho^{\textrm{LT}}$. We measure the percent increase in total mass needed to mimic the increase in velocity due to the various Szekeres anisotropic models at $r=2$ Mpc and summarize this in Table \ref{table1}. We find that the increase in mass needed to mimic the anisotropies is on the order of a few percent and increases nonlinearly with the anisotropic mass fraction. Doubling the anisotropic mass fraction from $5\%$ to $10\%$ increases the total mass fraction needed by a factor of nearly 4, while tripling it to $15\%$ increases the mass fraction needed by more than a factor of 8. This nonlinear behavior agrees well with the magnitude increases seen in the relative growth rate of density and velocity measurements, confirming a consistent picture for the physical impacts of anisotropies as modeled by the Szekeres metric.

\section{Conclusion}\label{conclusion}

With the rapid improvement of the precision and availability of observational data, relaxing traditional assumptions of homogeneity or isotropy in our models of structures and the universe in order to question the robustness of our interpretations will be a key test for cosmology and astrophysics in the coming decades. We extended our previous work in this area by considering the impact of anisotropies on physical measurements at single galaxy cluster scales. We employed the inhomogeneous and anisotropic Szekeres metric, an exact solution to Einstein's field equations, to model a realistic galaxy cluster in general relativity based on a modified NFW density profile. We quantified and compared the effects of a simple, but not trivial, dipole anisotropy on physical properties at the $5\%$, $10\%$, and $15\%$ levels relative to a reference LT (isotropic) model. To do this, we also introduced a consistent measurement of the velocity of the dust in the Szekeres models relative to the traditional $\Phi,_{t}$ measure in LT models.

Using volume averages over shells of thickness 10 kpc, we compared the impact of the anisotropy in the Szekeres models on the growth rate of density within the structure, the relative velocity of neighboring shells of dust, and the total velocity toward the origin in the collapsing structure to the reference LT model. We found increases in the growth rate of density and the clustering velocity of the dust over the LT model in anisotropic regions of the Szekeres structure, providing a consistent kinematic picture of increased growth due to anisotropies in a collapsing structure. The relative impact due to the anisotropies on these two measures, as well as the total velocity toward the center of the structure, increases nonlinearly, such that doubling the anisotropic mass fraction from $5\%$ to $10\%$ increases the effect on all three physical measures by a factor of nearly 4, while tripling it to $15\%$ increases the effect by a factor of 8 to 10.

We find at the $10\%$ level of anisotropy an average increase in the growth rate of density $\langle \rho,_{t}\rangle_{\mathcal{D}}-\langle \rho,_{t}^{\textrm{LT}}\rangle_{\mathcal{D}}$ of 20 to 40, an average increase in the relative velocity of shells of dust with separation 10 kpc of 14 km s${}^{-1}$, and an average increase in velocity toward the center of the structure of 10 km s${}^{-1}$. This is compared to the LT reference structure, which has a maximum infall velocity of nearly 500 km s${}^{-1}$, and represents a 2\% change in the maximum infall velocity. This increases to 20 km s${}^{-1}$ for a 15\% anisotropy, which is a 5\% increase relative to the LT structure. We have also shown that the impact of the anisotropy on the determinations of physical properties like the total mass would be impacted at the $1.0\%$, $3.8\%$ and $8.4\%$ levels for $5\%$, $10\%$, and $15\%$ levels of anisotropy in the structure.

This work represents a preliminary effort to construct a realistic, anisotropic Szekeres model of a cluster of galaxies following the traditional NFW dark matter halo description. The anisotropy represented is merely a dipole perturbation on this reference LT dark matter halo, and thus would represent small deviations from spherical symmetry in the halo in the form of halo substructure. The impact on physical properties like growth rate of density and velocity of the dust indicate that realistic levels of anisotropy could cause a strong bias on percent level measurements of the kinematic properties of clusters. The strongly nonlinear response to the level of anisotropy may indicate more substantial impacts on observables in general anisotropic models of structures that are not merely perturbations on a spherically symmetric halo, and further work is needed to model such anisotropies and determine the degree to which they will bias observable properties of the cluster.

\acknowledgments

We thank L. King and W. Rindler for useful comments during the preparation of this work. 
MI acknowledges that this material is based upon work supported in part by NSF under grant AST-1109667 and by NASA under grant NNX09AJ55G, and that part of the calculations for this work have been performed on the Cosmology Computer Cluster funded by the Hoblitzelle Foundation.
MT acknowledges that this work was supported in part by the NASA/TSGC Graduate Fellowship program.

\appendix

\section{Volume averaging in the Szekeres metric}\label{app1}

The volume integral over some domain $\mathcal{D}$ can be written 
\be	
\Psi_{|\mathcal{D}}=\int_{\mathcal{D}}d^3x\sqrt{\det{(g_{ij})}}\Psi.\label{eq:volint}
\ee
The volume average for some time $t$ of a scalar field $\Psi$ over some domain $\mathcal{D}$ is then
\be	
\langle \Psi\rangle_{\mathcal{D}}=\frac{1}{V_{\mathcal{D}}}\int_{\mathcal{D}}d^3x\sqrt{\det{(g_{ij})}}\Psi,\label{eq:volavg}
\ee
where $V_{\mathcal{D}}$ is the volume of the domain
\be	
V_{\mathcal{D}}=\int_{\mathcal{D}}d^3x\sqrt{\det{(g_{ij})}},\label{eq:vol}
\ee
and $g_{ij}$ is the 3D spatial part of the Szekeres metric
\bea	
ds^2&=&-dt^2+g_{ij}dx^idx^j\\
&=&-dt^2+\frac{H^2}{1-k(r)}dr^2+\Phi^2(d\theta^2+\sin^2\theta d\phi^2)\nonumber,
\eea
where $H\equiv \Phi,_{r}-\Phi\E,_{r}/\E$. We have rewritten the metric in Eq. (\ref{eq:metric}) in angular coordinates $\theta$ and $\phi$ \cite{coord} through the use of Eq. (\ref{eq:transform}), because analytical integration of quantities involving $\E,_{r}/\E$ is more straightforward in these coordinates. 

If we choose the domain $\mathcal{D}$ of the volume integral to be the shell between some $r_0$ to $r$, the volume average can then be written 
\be	
\langle \Psi\rangle_{\mathcal{D}}=\frac{1}{V_{\mathcal{D}}}\int_{r_0}^{r}dr\frac{\Phi^2}{\sqrt{1-k(r)}}\int\sin\theta d\theta\int d\phi H \Psi.\label{eq:volavgsz}
\ee
The full volume average interior to some radius $r$ then simply requires $r_0=0$. Noting that in $(\theta,\phi)$ coordinates
\be	
\frac{\E,_{r}}{\E}=-\frac{S,_{r}\cos\theta+(P,_{r}\cos\phi+Q,_{r}\sin\phi)\sin\theta}{S},
\ee
it is straightforward to show that
\be	
\int\sin\theta d\theta\int d\phi\frac{\E,_{r}}{\E}=0\label{eq:zero}
\ee
and
\bea
\int\sin\theta d\theta\int d\phi\left(\frac{\E,_{r}}{\E}\right)^2&=&\frac{4\pi}{3}\frac{S,_{r}^2+P,_{r}^2+Q,_{r}^2}{S^2}\\
&=&\frac{4\pi}{3}\left(\frac{\E,_{r}}{\E}\right)_{\textrm{ext}}^2,\label{eq:s2}
\eea
where the subscript 'ext' denotes the extreme value. The volume of such a domain is then just
\be	
V_{\mathcal{D}}=4\pi\int_{r_0}^{r}dr\frac{\Phi^2\Phi,_{r}}{\sqrt{1-k(r)}},
\ee
which is independent of $\E,_{r}/\E$.

The presence of a surface integral in the domain average like that in Eq. (\ref{eq:zero}) then determines whether the Szekeres dipole contribution from $\E,_{r}/\E$ will contribute to the averaged quantity. Any surface integral over a quantity in the volume average of Eq. (\ref{eq:volavgsz}) (after multiplication by the term $\Phi,_{r}-\Phi\E,_{r}/\E$) that contains only odd powers of $\E,_{r}/\E$ will have no contribution from the Szekeres dipole due to Eq. (\ref{eq:zero}) and behave identically to the corresponding LT metric (with $\E,_{r}/\E=0$). Alternately, any integral which contains at least one even power of $\E,_{r}/\E$ can still have a contribution from the Szekeres dipole.


\begin{thebibliography}{}

\bibitem{nfw97} J. F. Navarro, C. S. Frenk and S. D. M. White, \textit{A Universal Density Profile from Hierarchical Clustering}, \apj~\textbf{490} (1997) 493.
\bibitem{SKMHH2003} H. Stephani, D. Kramer, M. MacCallum, C. Hoenselaers and E. Herlt, \textit{Exact Solutions of Einstein's Field Equations}, Cambridge University Press, Cambridge U.K. (2003).
\bibitem{Krasinski1997} A. Krasi\'nski, \textit{Inhomogeneous Cosmological Models}, Cambridge University Press, Cambridge U.K. (1997).
\bibitem{Szekeres1} P. Szekeres, \textit{A class of inhomogeneous cosmological models}, Commun. Math. Phys. \textbf{41} (1975) 55.
\bibitem{Szekeres2} P. Szekeres, \textit{Quasispherical gravitational collapse}, Phys. Rev. D \textbf{12}, 2941 (1975).
\bibitem{Bonnor&Tomimura1976} W. Bonnor and N. Tomimura, \textit{Evolution of Szekeres's cosmological models}, Mon. Not. R. Astron. Soc. \textbf{175}, 85 (1976).
\bibitem{Ishaketal2008}  M. Ishak, J. Richardson, D. Garred, D. Whittington, A. Nwankwo and R. Sussman, \textit{Dark energy or apparent acceleration due to a relativistic cosmological model more complex than the Friedmann-Lemaitre-Robertson-Walker model?}, Phys. Rev. D \textbf{78}, 123531 (2008).
\bibitem{Nwankwoetal2011} A. Nwankwo, M. Ishak and  J. Thompson, \textit{Luminosity distance and redshift in the Szekeres inhomogeneous cosmological models}, J. Cosmol. Astropart. Phys. 05 (2011) 028. 
\bibitem{Bolejko&Celerier2010} K. Bolejko and M.-N. C\'el\'erier, \textit{Szekeres Swiss-cheese model and supernova observations}, Phys. Rev. D \textbf{82}, 103510.
\bibitem{Ishak&Peel2012} M. Ishak and A. Peel, \textit{Growth of structure in the Szekeres class-II inhomogeneous cosmological models and the matter-dominated era}, Phys. Rev. D \textbf{85}, 083502 (2012).
\bibitem{Peel2012} A. Peel, M. Ishak and M.A. Troxel, \textit{Large-scale growth evolution in the Szekeres inhomogeneous cosmological models with comparison to growth data}, Phys. Rev. D \textbf{86}, 123508 (2012).
\bibitem{BolejkoCMB} K. Bolejko, \textit{The Szekeres Swiss Cheese model and the CMB observations}, Gen. Rel. Grav. \textbf{41}, 1737-1755 (2009). 
\bibitem{Buckley} R. Buckley and E. Schlegel, \textit{CMB dipoles and other low-order multipoles in the quasispherical Szekeres model}, Phys. Rev. D \textbf{87}, 023524 (2013).
\bibitem{barrowsilk} J. D. Barrow and J. Silk, \textit{The growth of anisotropic structures in a Friedmann universe}, \apj \textbf{250}, 432 (1981).
\bibitem{kantowskisachs} R. Kantowski and R. K. Sachs, \textit{Some Spatially Homogeneous Anisotropic Relativistic Cosmological Models}, J. Math. Phys. \textbf{7}, 443 (1966).

\bibitem{Bonnoretal1977} W. Bonnor, A. H. Sulaiman and N. Tomimura, \textit{Szekeres's space-times have no Killing vectors}, Gen. Relativ. Gravit. \textbf{8}, 549 (1977).
\bibitem{Ellis&VanElst1998} G. F. R. Ellis and H. van Elst, Cosmological Models (Carg\`ese Lectures 1998). In \textit{Theoretical and Observational Cosmology}, Ed. M. Lachieze-Rey. (Kluwer, Nato Series C: Mathematical and Physical Sciences, Vol. 541, 1999), p. 1-116.
\bibitem{Bolejko2006} K. Bolejko, \textit{Structure formation in the quasispherical Szekeres model}, Phys. Rev. D \textbf{73}, 123508 (2006).

\bibitem{Jing2002} Y. P. Jing and Y. Suto, \textit{Triaxial Modeling of Halo Density Profiles with High-Resolution N-Body Simulations}, \apj \textbf{574}, 538 (2002).
\bibitem{Oguri2003} M. Oguri, J. Lee and Y. Suto, \textit{Arc Statistics in Triaxial Dark Matter Halos: Testing the Collisionless Cold Dark Matter Paradigm}, \apj \textbf{599}, 7 (2002).
\bibitem{Tormen1998} G. Tormen, A. Diaferio and D. Syer, \textit{Survival of substructure within dark matter haloes}, \mnras \textbf{299}, 728 (1998).
\bibitem{Sheth2003} R. K. Sheth and B. Jain, \textit{Substructure and the halo model of large-scale structure}, \mnras~\textbf{345}, 529 (2003).
\bibitem{Giocoli2010} C. Giocoli, M. Bartelmann, R. K. Sheth and M. Cacciato, \textit{Halo model description of the non-linear dark matter power spectrum at $k >> 1$Mpc${}^{-1}$}, \mnras~\textbf{408}, 300 (2010).
\bibitem{Oguri2004} M. Oguri and J. Lee, \textit{A realistic model for spatial and mass distributions of dark halo substructures: An analytic approach}, \mnras \textbf{355}, 120 (2004).
\bibitem{Corless2007} V. Corless and L. King, \textit{A statistical study of weak lensing by triaxial dark matter haloes: consequences for parameter estimation}, \mnras~\textbf{380}, 149 (2007).
\bibitem{Bahe2012} Y. M. Bahe, I. G. McCarthy and L. King, \textit{Mock weak lensing analysis of simulated galaxy clusters: bias and scatter in mass and concentration}, \mnras \textbf{421}, 1073 (2012).

\bibitem{Lemaitre1933} G. Lema\^itre, \textit{L'Univers en expansion}, Ann. Soc. Sci. Bruxelles \textbf{A53}, 51 (1933).
\bibitem{tolman1934} R. C. Tolman, \textit{Effect of Inhomogeneity on Cosmological Models}, Proc. Nat. Acad. Sci. USA \textbf{20}, 169 (1934).

\bibitem{Hellaby1996} C. Hellaby, \textit{The null and KS limits of the Szekeres model}, Class. Q. Grav. \textbf{13}, 2537 (1996).
\bibitem{agenote} We choose $t_0$ in this way despite setting $\Lambda=0$ in the model, because we want to be consistent in the model's ability to match at $t_0$ to a standard FLRW cosmology outside of the structure. The higher $t_0$ also prevents the curvature $k(r)$ from becoming hyperbolic ($k(r)<1$) at higher $r$-values, which is beneficial in the numerical evaluation of $\Phi$.
\bibitem{krasinski2002} A. Krasi\'nski and C. Hellaby, \textit{Structure formation in the Lemaître-Tolman model}, Phys. Rev. D \textbf{65} 023501 (2002).
\bibitem{hellaby2002} C. Hellaby and A. Krasi\'nski, \textit{You cannot get through Szekeres wormholes: Regularity, topology, and causality in quasispherical Szekeres models}, Phys. Rev. D \textbf{66}, 084011 (2002).
\bibitem{hellaby2006} C. Hellaby and A. Krasi\'nski, \textit{Alternative methods of describing structure formation in the Lemaitre-Tolman model}, Phys. Rev. D \textbf{73}, 023518 (2006).
\bibitem{walters2012} A. Walters and C. Hellaby, \textit{Constructing realistic Szekeres models from initial and final data}, J. Cosmol. Astropart. Phys. \textbf{12}, 001 (2012).

\bibitem{baltz2009} E. A. Baltz, P. Marshall and M. Oguri, \textit{Analytic models of plausible gravitational lens potentials}, J. Cosmol. Astropart. Phys. \textbf{1}, 15 (2009).

\bibitem{lcdm} The space outside the structure is not $\Lambda$CDM due to our choice of $\Lambda=0$, which makes the numerical calculation of $\Phi$ simpler. We are only interested in the dynamics of the galaxy cluster, where $\Lambda$ is negligible, and do not consider the evolution of the universe outside of the structure beyond ensuring there are no shell-crossing singularities. The model could be matched to a $\Lambda$CDM space at $t_0$ by instead including a $\Lambda\ne0$ and recalculating the function $\Phi$, but this is not needed for our purpose here. 

\bibitem{Plebanski&Krasinski2006} J. Pleba\'nski and A. Krasi\'nski, \textit{An Introduction to General Relativity and Cosmology}, Cambridge University Press, Cambridge U.K. (2006).
\bibitem{bolejko2009} K. Bolejko, \textit{Volume averaging in the quasispherical Szekeres model}, Gen. Rel. Grav. \textbf{41}, 1585 (2009).
\bibitem{sussman2012} R. A. Sussman and K. Bolejko, \textit{A novel approach to the dynamics of Szekeres dust models}, Class. Q. Grav. \textbf{29}, 065018 (2012).
\bibitem{munari2013} M. E. Munari, A. Biviano, S. Borgani, G. Murante and D. Fabjan, \textit{The relation between velocity dispersion and mass in simulated clusters of galaxies: dependence on the tracer and the baryonic physics}, \mnras \textbf{430}, 2638 (2013).
\bibitem{coord} These coordinates are not the usual spherically symmetric coordinates, as the $r$ coordinate is not radial in the Szekeres metric.

\end{thebibliography}
\end{document}